\documentclass[10pt, journal]{IEEEtran}
\ifCLASSINFOpdf
\else
\fi
\usepackage{amsfonts}
\usepackage{amsmath}
\usepackage{amsthm}
\usepackage{amssymb}
\usepackage{amsfonts}
\usepackage{amsmath}
\usepackage{amsthm}
\usepackage{graphicx}
\usepackage{epstopdf}
\usepackage{cite}
\newtheorem*{theo}{Theorem}

\begin{document}
\title{ Optimal DoF region of the $K$-User MISO BC with Partial CSIT}

\author{ \IEEEauthorblockN{Enrico~Piovano
		and Bruno~Clerckx,~\IEEEmembership{Senior Member,~IEEE}} \\
	    \thanks{E. Piovano and B. Clerckx are with the Department of Electrical and Electronic Engineering, Imperial College London, SW7 2AZ, UK (e-mail: \{e.piovano15, b.clerckx\}@imperial.ac.uk). }
	   \thanks{This work has been partially supported by the EPSRC of UK, under grant EP/N015312/1.} }
\maketitle
\begin{abstract}	
We consider the $K$-User Multiple-Input-Single-Output (MISO) Broadcast Channel (BC) where the transmitter, equipped with $M$ antennas, serves $K$ users, with $K \leq M$.
The transmitter has access to a partial 
channel state information of the users. This is modelled by letting the variance of the Channel State Information at the Transmitter (CSIT) error of user $i$ scale as $O(P^{-\alpha_i}$) for the Signal-to-Noise Ratio (SNR) $P$ and some constant $\alpha_i \geq 0$.
In this work we derive the optimal Degrees-of-Freedom (DoF) region in such setting and we show that Rate-Splitting (RS) is the key scheme to achieve such a region.

\end{abstract}
\IEEEpeerreviewmaketitle
\section{Introduction} \label{Introduction}
The use of multiple antennas at the transmitter has dramatically increased the capacity of wireless networks, as multiple antennas can help to achieve a larger number of Degrees-of-Freedom (DoF).
However, in order to achieve the theoretical multiplexing gain,  a sufficiently accurate Channel State Information at the Transmitter (CSIT) is required \cite{Jindal2006,Caire2010}.  Nonetheless, acquiring an accurate CSIT is a difficult task.
 
 In this paper we investigate the DoF region of the  \mbox{$K$-User} Multiple-Input-Single-Output (MISO) Broadcast Channel (BC), where the transmitter has a partial knowledge of the channel of the users.
 As in  \cite{Davoodi2016,Yang2013}, the partial CSIT is captured by letting the variance of the channel estimation error of user $i$ decay as $O(P^{-\alpha_i})$ for some exponent $\alpha_i \in [0,1]$, which represents the CSIT quality.  
 Under such setting, a great deal of research has mostly focused on \mbox{characterizing the Sum-DoF}.
 A key result was shown in \cite{Davoodi2016}, where by assuming that the CSIT qualities of the users are arranged as $\alpha_1 \geq \dots \geq \alpha_K$, it was proved that a Sum-DoF upperbound is given by $1+\alpha_2+\dots+\alpha_K$.
 Moreover, such an upperbound is achievable through a Rate-Splitting (RS) strategy \cite{Hao2015,Joudeh2016,Joudeh2016a}. 
 While the Sum-DoF is an important information to know, it does not reveal any information about the
 individual DoF achieved by each user but only about the sum.
The individual DoF of the users are instead characterized by the DoF region, which is the set of all achievable DoF tuples $(d_1, \dots, d_K)$.
However, since taking into consideration the DoF achieved by each user is difficult, to describe the DoF region is a challenging task.

 In this work, to the best of our knowledge, we characterize for the first time the optimal DoF region in the above setting. 
 Building upon the work in \cite{Davoodi2016}, we derive an outer-bound of the optimal region, which is a polyhedron.
 We then show the achievability of such an outer-bound, which is the main challenge of this work since we have to show the achievability of each tuple $(d_1, \dots, d_K)$ and not just the achievability of the sum. 
 We introduce an original approach: instead of characterizing and showing the achievability of the corner points of the polyhedron which looks unfeasible for a large number of users, we characterize and show the achievability of each facet of the polyhedron.  
 The key strategy for the achievability is RS with flexible power allocation.
 Hence, RS is not only optimal to achieve the \mbox{Sum-DoF}, \mbox{but also to achieve the DoF region.} 
\section{System Model} \label{system_model}
This work considers a setup where a transmitter, equipped with $M$ antennas, serves $K$ single-antenna users, with  \mbox{$K \leq M$}.
The users are indexed by the set $\mathcal{K}=\{1,\ldots,K\}$.
At $t$-th channel use, the signal received by the \mbox{$i$-th} receiver is
\begin{equation} \label{rx_signal}
y_i(t)=\mathbf{h}^H_i(t)\mathbf{x}(t)+n_i(t)
\end{equation}
where $\mathbf{h}_i^H(t) \in \mathbb{C}^{1 \times M}$  is the  channel vector and
$\mathbf{x}(t) \in \mathbb{C}^{M \times 1}$ is the transmitted signal, which is subject to the power constraint $\mathbb{E}(\| \mathbf{x}(t) \|^2) \leq P $. 
The term $n_i(t) \sim \mathcal{CN}(0,1)$ indicates the additive noise.
We define the channel matrix $\mathbf{H}(t)=\left[\mathbf{h}_1(t),\dots,\mathbf{h}_K(t) \right]^H \in \mathbb{C}^{K \times M} $, drawn from a continuous ergodic distribution and such that the joint density of its elements exists. We assume that the matrix and all its sub-matrices are full-rank. 
In addition, to avoid degenerate situations, we assume that the entries and the determinant of $\mathbf{H}(t)$ are bounded away from zero and infinity \cite{Davoodi2016}.

For each user $i$, the transmitter has a current estimate of the channel, indicated as $\hat{\mathbf{h}}_i(t)$. 
The partial CSIT is modelled as \mbox{${\mathbf{h}}_i(t)={\hat{\mathbf{h}}}_i(t)+\tilde{\mathbf{h}}_i(t)$}, where $\tilde{\mathbf{h}}_i(t)$ is the channel estimation error at the transmitter.
${\hat{\mathbf{h}}}_i(t)$ and 
${\tilde{\mathbf{h}}}_i(t)$  are assumed to be uncorrelated.
Furthermore, the CSIT error ${\tilde{\mathbf{h}}}_i(t)$ has i.i.d. entries $\mathcal{CN}(0,\sigma_i^2)$,  where $\sigma_i^2 \leq 1$,  while the entries of ${\hat{\mathbf{h}}}_i(t)$ have a variance equal to $1-\sigma_i^2$.
For the sake of notational convenience, the index $t$ of the channel use is omitted in the rest of the paper.
The variance $\sigma_i^2$ is assumed to decay with the SNR $P$ as $O(P^{-\alpha_i})$, where $\alpha_i$ is defined as the  \textit{CSIT quality exponent}. 
We can restrict the exponent $\alpha_i$ to the case $\alpha_i \in [0,1]$ since, from a DoF perspective, $\alpha_i=0$ offers no gain over a no CSIT case 
while  $\alpha_i \geq 1$ corresponds to a perfect CSIT.
We assume, without loss of generality, that users are ordered with respect to their CSIT quality, i.e.  $\alpha_1 \geq \alpha_2 \geq \ldots \geq \alpha_K$.
We also remind that, given a unitary Zero-Forcing (ZF) precoded vector $\mathbf{v}$ such that $\hat{\mathbf{h}}_i^H\mathbf{v}=0$, the equation $\mathbb{E}[|{\mathbf{h}}_i^H\mathbf{v}|^2]=O(P^{-\alpha_i})$ is satisfied.

The transmitter has messages $W_1,\ldots, W_K$ intended for the corresponding users.
Codebooks, probability of error, achievable rate tuples $(R_1(P),\ldots,R_K(P))$ and the capacity region $\mathcal{C}(P)$ are all defined in the Shannon theoretic sense.
The DoF tuple $(d_1, \ldots, d_K)$ is said to be achievable if there exists $(R_1(P),\ldots,R_K(P)) \in \mathcal{C}(P)$ such that $d_i=\lim_{P \to \infty} \frac{R_i(P)}{\log(P)}$ for all $i \in \mathcal{K}$.
The DoF region is defined as the closure of all achievable DoF tuples $(d_1,\dots, d_K)$ and is denoted by $\mathcal{D}^*$.
\section{Main Result} \label{DoF region}
In order to state the main result of the paper, we define $\mathcal{A}$ as the set of all possible non-empty subsets of $\mathcal{K}$ with elements arranged in an ascending order.
For instance, in case of $\mathcal{K}=\{1,2,3\}$, the set $\mathcal{A}$ is given by \mbox{$\mathcal{A}=\{ \{1\},\{2\},\{3\}, \{1,2\}, \{1,3\}, \{2,3\}, \{1,2,3\} \}$}.
Any element of $\mathcal{A}$, which is itself a set, is indicated with a calligraphic upper case letter and its elements are denoted with the corresponding lower case letter (with numbered subscripts). For instance  \mbox{$\mathcal{S}=\{s_1,s_2,\ldots,s_{|\mathcal{S}|}\} \in \mathcal{A}$}, 
where \mbox{$s_1 < s_2 < \ldots < s_{|\mathcal{S}|}$}.
The main result is the following.
\begin{theo} \label{theo_proposed_scheme}
The optimal DoF region $\mathcal{D}^*$ of the $K$-User MISO BC with partial CSIT is given by  
all the real tuples $(d_1,\ldots,d_K)$ which satisfy
\begin{equation} \label{dis_1}
  	 d_i \geq 0, \quad \forall i \in \mathcal{K}
  	\end{equation}
	\begin{equation} \label{dis_2}
	\sum_{i \in \mathcal{S}}{d_{i}}  \leq 1 + \sum_{i \in \mathcal{S} \setminus \{s_1\}}{\alpha_i}, \quad \forall \mathcal{S} \in \mathcal{A}.
	\end{equation}
\end{theo}
We denote as $\mathcal{D}$ the above region described by the inequalities (\ref{dis_1}) and (\ref{dis_2}). In order to show that $\mathcal{D}$ coincides with the optimal DoF region $\mathcal{D}^*$, we need to show that $\mathcal{D}$ is simultaneously an outer-bound of the optimal region and is achievable.
The fact that $\mathcal{D}$ is an outer-bound of $\mathcal{D}^*$ follows after few steps from \cite[Th. 1]{Davoodi2016}, which states that the Sum-DoF of the $K$-User MISO BC, with $K \leq M$, is upperbounded by 
\begin{equation} \label{upper-bound_Jafar} 
\sum_{i \in \mathcal{K}}{d_{i}}  \leq 1 + \sum_{i \in \mathcal{K} \setminus \{1\}}{\alpha_i}.
\end{equation}
The result was shown assumining $\alpha_1=1$ for the first user.
However, since enhancing  the CSIT  does not harm the \mbox{Sum-DoF}, the same upperbound holds for a generic value of $\alpha_1 \in [0,1]$. 
The region $\mathcal{D}$ is constructed by applying such a Sum-DoF upperbound to any arbitrary subset of users $\mathcal{S} \in \mathcal{A}$, which states that the Sum-DoF of users in $\mathcal{S}$ is upperbounded by
$\sum_{i \in \mathcal{S} }{d_i} \leq 1 + \sum_{i \in \mathcal{S} \setminus \{s_1\}}{\alpha_i}$.
Considering all possible subsets of users $\mathcal{S} \in \mathcal{A}$ and given 
that the DoF of each user is a non-negative real value, 
we obtain $\mathcal{D}$ as an outer-bound of the optimal DoF region $\mathcal{D}^*$.
The challenge of the paper is to show the achievability of $\mathcal{D}$, addressed in Section \ref{proof}. This means to show that each DoF tuple $(d_1,\ldots,d_K)$ of $\mathcal{D}$, which takes into consideration the individual DoF achieved by each user and not just the sum, is achievable.
\section{Rate-Splitting scheme} \label{RS_scheme}
In this section, we remind the RS scheme which will be used to show the achievability of $\mathcal{D}$ in Section \ref{proof}.
In RS, we transmit two kinds of symbols that are superimposed in the power domain: a common symbol decoded by all users on top of private symbols decoded by the respective users only.
This strategy has been shown to be more robust in treating interference when partial CSIT is available compared to conventional linear precoding schemes (where only private symbols are transmitted) \cite{Yang2013, Hao2015, Joudeh2016}.
Getting into the details of the scheme, the message of each user $i \in \mathcal{K}$ is split into \mbox{$W_i=(W_i^{(\mathrm{c})},W_i^{(\mathrm{p})})$}, where $W_i^{(\mathrm{c})}$ is a common (or public) sub-message  while $W_i^{(\mathrm{p})}$ is a private sub-message. 
All the common sub-messages $W_1^{(\mathrm{c})},\ldots,W_K^{(\mathrm{c})}$ are jointly encoded into the common symbol $x^{\mathrm{(c)}}$, which has to be decoded by all $K$ users. 
Each private sub-message $W_i^{(\mathrm{p})}$ is encoded into the private symbol $x_i^{(\mathrm{p})}$, which is decoded by user $i$ only. 
It is assumed that all the symbols are drawn from a unitary-power Gaussian codebook. 
Next, the symbols are linearly precoded and power allocated.
The transmitted signal takes the form 
\begin{equation} \label{tx_phase_1}
\mathbf{x}=\sqrt{P^{(\mathrm{c})}}\mathbf{v}^{(\mathrm{c})}x^{(\mathrm{c})}+\sum_{i \in \mathcal{K}}{\sqrt{P_i^{(\mathrm{p})}}\mathbf{v}_i^{(\mathrm{p})}x_i^{(\mathrm{p})}}
\end{equation}
where $\mathbf{v}^{(\mathrm{c})} \in \mathbb{C}^{M \times 1}$ and $\mathbf{v}_i^{(\mathrm{p})} \in \mathbb{C}^{M \times 1}$ are unitary precoding vectors, and $P^{(\mathrm{c})}$ and $P_i^{(\mathrm{p})}$ are the corresponding allocated powers with $P^{(\mathrm{c})}+\sum_{i \in \mathcal{K}}{P_i^{(\mathrm{p})}} \leq P$.
Since the common symbol has to be decoded by all users, $\mathbf{v}^{(\mathrm{c})}$ is chosen as a random (or generic) precoding vector.
On the other hand, the private symbols are precoded by ZF over the channel estimate, i.e.
$\mathbf{v}_i^{(\mathrm{p})} \perp \big\{\hat{\mathbf{h}}_l \big\}_{l \in \mathcal{K} \setminus \{i\}}$.
The power allocation is set such that $P^{(\mathrm{c})} = O(P)$ and $P_i^{(\mathrm{p})} = O(P^{a_i})$, where $a_i$ correspond to the power levels and are such that $a_i \in [0,1]$. The values of $a_i$ are concatenated into the vector $(a_1, \ldots, a_K)$. 

The received signal in (\ref{rx_signal}) for user $j \in \mathcal{K}$ is given by
\begin{equation*} 
y_j=\underbrace{\sqrt{P^{(\mathrm{c})}}\mathbf{h}_j^H\mathbf{v}^{(c)}x^{(\mathrm{c})}}_{O(P)}+
\underbrace{\sqrt{P_j^{(\mathrm{p})}}\mathbf{h}_j^H\mathbf{v}_j^{(\mathrm{p})}x_j^{(\mathrm{p})}}_{O(P^{a_j})}
\end{equation*}
\begin{equation} \label{rx_RS_signal}
 + \sum_{i \in \mathcal{K} \setminus \{j\}}{\underbrace{\sqrt{P_i^{(\mathrm{p})}}\mathbf{h}_j^H \mathbf{v}_i^{(\mathrm{p})}x_i^{(\mathrm{p})}}_{O(P^{a_i-\alpha_j})}} + \underbrace{n_j}_{O(1)}.
\end{equation}
All users decode the common symbol by treating the interference from all other private symbols as noise.
From  (\ref{rx_RS_signal}), it can be verified that the common symbol $x^{(\mathrm{c})}$, in order to be successfully decoded by all users, can carry a DoF of 
\begin{equation} \label{DoF_common_symbol} 
d^{(\mathrm{c})} = 1-\max_{j \in \mathcal{K}} a_j.
\end{equation}
The DoF of the common symbol can be split in all possible ways among users in $\mathcal{K}$.
We denote as $d_j^{(\mathrm{c})}$ the DoF of the common symbol given to user $j$.
It follows that any non-negative real tuple $(d_1^{(\mathrm{c})},\ldots, d_K^{(\mathrm{c})})$, which satisfies $\sum_{j \in \mathcal{K}}{d_j^{(\mathrm{c})}} = d^{(\mathrm{c})}$, is an admissible partition of the DoF carried by the common symbol
among the users in $\mathcal{K}$.

Next, each user removes $x^{(\mathrm{c})}$ by performing Successive Interference Cancellation (SIC) and proceeds to decode its own private symbol. 
From (\ref{rx_RS_signal}), the private symbol intended for user $j \in \mathcal{K}$ can carry a DoF of 
\begin{equation} \label{DoF_private_RS} 
d_j^{(\mathrm{p})} = \left( a_j -\left( \max_{i \in \mathcal{K} \setminus \{j\}}{a_i} - \alpha_j \right)^+ \right)^{+}.
\end{equation}
where $(x)^+=\max\{x,0\}$.
The DoF of all the private symbols are collected into the vector
$(d_1^{(\mathrm{p})}, \ldots, d_K^{(\mathrm{p})})$.
To sum up, a DoF tuple $\left(d_1,\ldots,d_K \right)$ is achievable by RS with power levels given by $\left(a_1,\ldots,a_K \right)$ if the following equality holds:
\begin{equation} \label{achievable_DoF} 
(d_1, \ldots, d_K)=(d_1^{(\mathrm{p})}, \ldots, d_K^{(\mathrm{p})})+(d_1^{(\mathrm{c})}, \ldots, d_K^{(\mathrm{c})})
\end{equation}
where $d_j^{(\mathrm{p})}$ for any $j \in \mathcal{K}$ is given
by (\ref{DoF_private_RS}), while $(d_1^{(\mathrm{c})}, \ldots, d_K^{(\mathrm{c})})$ indicates an admissible partition of the total DoF carried by the common symbol, which is given by (\ref{DoF_common_symbol}), as described above. 

RS outperforms conventional linear precoding scheme, as Zero-Forcing Beamforming (ZFBF), in case of partial CSIT. In particular RS attains the Sum-DoF
 upperbound in (\ref{upper-bound_Jafar}), which is achievable
considering $(a_j)_{j \in \mathcal{K}}=b$, for any $b$ such that $\alpha_2 \leq b \leq \alpha_1$, and any split of the DoF carried by the common symbol (which is irrelevant to the Sum-DoF). 
In fact, \mbox{from (\ref{DoF_private_RS})}, we have that such power allocation leads to \mbox{$d_1^{(\mathrm{p})}=b$} and \mbox{$d_j^{(\mathrm{p})}=\alpha_j$} for \mbox{$j \in \mathcal{K} \setminus \{1\}$}, while  $d^{(\mathrm{c})}=1-b$ from (\ref{DoF_common_symbol}). 
Hence, the Sum-DoF is equal \mbox{to (\ref{upper-bound_Jafar})}. 
It is important to notice that ZFBF achieves a Sum-DoF of $\alpha_1+\ldots+\alpha_K$. Hence, it only attains the upperbound in (\ref{upper-bound_Jafar}) for $\alpha_1=1$, while it fails when $\alpha_1 < 1$, where RS is needed.
\section{Proof of the achievability of $\mathcal{D}$} \label{proof}
In this section we show the achievability of $\mathcal{D}$ characterized in Section \ref{DoF region}. 
The region $\mathcal{D}$ is the $K$-dimensional polyhedron given by the intersection of the half-spaces described by (\ref{dis_1}) and (\ref{dis_2}).
We show that $\mathcal{D}$ is achievable by induction over the number of users $K$, considering
a number of antennas at the transmitter $M \geq K$. 
The hypothesis is clearly true for $K=1$. We assume that the hypothesis is valid for \mbox{$K=1,\ldots,k-1$} and we consider
the case $K=k$.
First, the half-spaces in (\ref{dis_1}) and (\ref{dis_2}) are delimited by the hyperplanes obtained by substituting the half-spaces' inequalities with equalities.
In total, there are $2^K+K-1$ hyperplanes.
Any of these hyperplanes contains a facet of the polyhedron $\mathcal{D}$ and the set of all the facets corresponds to the boundary of $\mathcal{D}$.
In our paper we show the achievability of $\mathcal{D}$ in a novel way. Instead of characterizing and 
showing the achievability of the corner points as in \cite{Yang2013}, we show the achievability of 
$\mathcal{D}$ by characterizing and showing the achievability of each of its facets. 
In fact, in \cite{Yang2013}, only the two-user case was considered. In such a case the two dimensional region boils down 
to a polygon and the corner points are simple to characterize.
However, the characterization of the corner points looks unfeasible for the $K$-dimensional case. 
Since a corner point is given by the intersection of $K$ hyperplanes, characterizing the corner points means to analyse each of the $\binom{2^K+K-1}{K}$ subsets of $K$ hyperplanes to see if they intersect in a point. When a subset of $K$ hyperplanes intersects in a point, we need to further verify if such a point belongs to the outer-bound. If the point belongs to the outer-bound, it is a corner point. Such 
procedure is unfeasible for large $K$.
Here, instead of finding the corner points, we propose a new approach where the facet contained in each of the hyperplanes delimiting
$\mathcal{D}$ is first characterized and then the achievability of each point of the facet is shown.
The facets from (\ref{dis_2}) will be shown to be achievable by RS with flexible power
allocation and flexible split of the common symbol, while the facets 
from (\ref{dis_1}) will be shown to be achievable by induction hypothesis. 
We first show the achievability of the facets contained in the 
hyperplanes which delimit the half-spaces in (\ref{dis_2}).
Any of these hyperplanes is given by $\sum_{i \in \mathcal{S}}{d_{i}}  =  1 + \sum_{i \in \mathcal{S} \setminus \{s_1\}}{\alpha_i}$, for a  subset $\mathcal{S} \in \mathcal{A}$. We denote the facet
contained in such an hyperplane as $\mathcal{F}_{\mathcal{S}}$.
The facet $\mathcal{F}_{\mathcal{S}}$ can be analytically characterized as the set of all the points contained in the hyperplane which
satisfy all the other inequalities of the polyhedron in (\ref{dis_1}) and (\ref{dis_2}). Hence, $\mathcal{F}_{\mathcal{S}}$ is the set of all non-negative real tuples $(d_1,\ldots,d_k)$ such that
\begin{equation} \label{ineq_G} 
\sum_{i \in \mathcal{G}}{d_{i}}  \leq 1 + \sum_{i \in \mathcal{G} \setminus \{g_1\}}{\alpha_i}, \quad  \forall {\mathcal{G}} \in {\mathcal{A}}, \; \mathcal{G} \neq \mathcal{S}
\end{equation}
\begin{equation} \label{ineq_S} 
\sum_{i \in \mathcal{S}}{d_{i}}  =  1 + \sum_{i \in \mathcal{S} \setminus \{s_1\}}{\alpha_i}
\end{equation}
where the elements of $\mathcal{G}$ (arranged in an increasing order) are indicated as $\mathcal{G}=\{g_1,\ldots,g_{|\mathcal{G}|}\}$.
While the inequalities in (\ref{dis_1}) are satisfied by considering non-negative real tuples, (\ref{ineq_S}) identifies the hyperplane containing $\mathcal{F}_{\mathcal{S}}$ and the inequalities in (\ref{ineq_G}) 
 identify all the other inequalities of $\mathcal{D}$ in (\ref{dis_2}).

Showing directly the achievability of
 $\mathcal{F}_{\mathcal{S}}$ by (\ref{ineq_G}) and (\ref{ineq_S})
 is a difficult task. 
 We start by rewriting $\mathcal{F}_{\mathcal{S}}$ in an equivalent form where the values which can be taken by $d_j$, for each user $j \in \mathcal{K}$, are bounded through inequalities.
This is obtained, for each $j \in \mathcal{K}$, by comparing 
 an inequality in (\ref{ineq_G}), considering a specific $\mathcal{G}$, with the equality in (\ref{ineq_S}).
Then we show that the new form of $\mathcal{F}_{\mathcal{S}}$ is achievable by RS.
We first consider the case $|\mathcal{S}| \geq 2 $.
We start by analysing the elements $j \in \mathcal{S}$.
In case of $j=s_1$, we consider the inequality in (\ref{ineq_G}) for the specific \mbox{$\mathcal{G}=\mathcal{S} \setminus \{s_1\}$} and the \mbox{equality in (\ref{ineq_S}), i.e.}
\begin{equation} 
\begin{cases}
\sum_{i \in \mathcal{S} \setminus \{s_1\}}{d_{i}}  \leq 1 + \sum_{i \in \mathcal{S} \setminus \{s_1,s_2\}}{\alpha_i} \\
\sum_{i \in \mathcal{S}}{d_{i}}  =  1 + \sum_{i \in \mathcal{S} \setminus \{s_1\}}{\alpha_i}.
\end{cases}
\end{equation}
By comparing the inequality and the equality, it follows that $d_{s_1} \geq \alpha_{s_2}$.
We then move to the case $j \in \mathcal{S} \setminus \{s_1\}$.
Here, we consider the inequality in (\ref{ineq_G}) for \mbox{$\mathcal{G}=\mathcal{S} \setminus \{j\}$}  and (\ref{ineq_S}), i.e.
\begin{equation} 
\begin{cases}
\sum_{i \in \mathcal{S} \setminus \{j\}}{d_{i}}  \leq 1 + \sum_{i \in \mathcal{S} \setminus \{s_1,j\}}{\alpha_i} \\
\sum_{i \in \mathcal{S}}{d_{i}}  =  1 + \sum_{i \in \mathcal{S} \setminus \{s_1\}}{\alpha_i}.
\end{cases}
\end{equation}
By comparison, it follows that $d_j \geq \alpha_j$.
Summarizing, for $j \in \mathcal{S}$, we have $d_{s_1} \geq \alpha_{s_2}$ and $d_{j} \geq \alpha_j$ for $j \in \mathcal{S} \setminus \{s_1\}$. 
Next, we analyse the elements \mbox{$j \in \bar{\mathcal{S}}$}, where \mbox{$\bar{\mathcal{S}}=\mathcal{K} \setminus \mathcal{S}$}. The set $\bar{\mathcal{S}}$ is partitioned into three subsets, denoted as $\bar{\mathcal{S}}_1$, $\bar{\mathcal{S}}_2$ and $\bar{\mathcal{S}}_3$, such that the subset
\mbox{$\bar{\mathcal{S}}_1=\{\, j \in \bar{\mathcal{S}} \mid j < s_1\,\}$}, the subset \mbox{$\bar{\mathcal{S}}_2=\{\, j \in \bar{\mathcal{S}} \mid s_1 < j < s_2\,\}$} and \mbox{$\bar{\mathcal{S}}_3=\{\, j \in \bar{\mathcal{S}} \mid j > s_2\,\}$}. 
In case of $j\in\bar{\mathcal{S}}_1$, we first compare the inequality in (\ref{ineq_G}) for the case \mbox{$\mathcal{G}=\mathcal{S} \cup \{j\}$} and the equality in (\ref{ineq_S}), i.e.
\begin{equation} 
\begin{cases}
\sum_{i \in \mathcal{S} \cup \{j\}}{d_{i}}  \leq 1 + \sum_{i \in \mathcal{S}}{\alpha_i} \\
\sum_{i \in \mathcal{S}}{d_{i}}  =  1 + \sum_{i \in \mathcal{S} \setminus \{s_1\}}{\alpha_i}.
\end{cases}
\end{equation}
It follows that \mbox{$d_j \leq \alpha_{s_1}$}.
We then compare the inequality (\ref{ineq_G}) for \mbox{$\mathcal{G}= (\mathcal{S}\cup \{j\} )  \setminus \{s_1\}$}   and the equality in (\ref{ineq_S}), i.e.
\begin{equation} 
\begin{cases}
\sum_{i \in (\mathcal{S}\cup \{j\} )  \setminus \{s_1\}}  {d_i} \leq 1 + \sum_{i \in \mathcal{S} \setminus \{s_1\}}{\alpha_i} \\
\sum_{i \in \mathcal{S}}{d_{i}}  =  1 + \sum_{i \in \mathcal{S} \setminus \{s_1\}}{\alpha_i}.
\end{cases}
\end{equation}
  It follows that \mbox{$d_j \leq d_{s_1}$}. Hence, $d_j \leq \min (\alpha_{s_1},d_{s_1})$ for \mbox{$j\in\bar{\mathcal{S}}_1$}.
	We then move to the case \mbox{$j \in \bar{\mathcal{S}}_2$}.
Proceeding as above, by comparing (\ref{ineq_G}) for the case \mbox{$\mathcal{G}= \mathcal{S} \cup \{j\}$} and (\ref{ineq_S}), we obtain $d_j \leq \alpha_{j}$.
Also, from (\ref{ineq_G}) for \mbox{$\mathcal{G}= (\mathcal{S}\cup \{j\} )  \setminus \{s_1\}  $}
and (\ref{ineq_S}), we obtain $d_j \leq d_{s_1}$. Hence, $d_j \leq \min (\alpha_j,d_{s_1})$ for $j\in\bar{\mathcal{S}}_2$. 
    Lastly, we consider $j \in \bar{\mathcal{S}}_3$. 
    By simply comparing (\ref{ineq_G}) for  \mbox{$\mathcal{G}= \mathcal{S} \cup \{j\}$} with
	 (\ref{ineq_S}), we get $d_j \leq \alpha_{j}$ for $j \in \bar{\mathcal{S}}_3$. 

	We can conclude that the facet $\mathcal{F}_{\mathcal{S}}$  is included in the set of all the non-negative real tuples $(d_1,\ldots,d_k)$ given by
	\begin{equation} \label{facet_S_outerbound_} 
	\begin{cases}
	d_{s_1} \geq \alpha_{s_2}\\
	d_j \geq \alpha_j, & j \in \mathcal{S} \setminus \{s_1\} \\
	d_j \leq \min(\alpha_{s_1},d_{s_1}), & j \in \bar{\mathcal{S}}_1 \\
	d_j \leq \min(\alpha_j,d_{s_1}), & j \in \bar{\mathcal{S}}_2 \\
	d_j \leq \alpha_j, & j \in \bar{\mathcal{S}}_3 \\
	\sum_{j \in \mathcal{S}}{d_{j}}  =  1 + \sum_{j \in \mathcal{S} \setminus \{s_1\}}{\alpha_j}.
	\end{cases}
	\end{equation}
	Furthermore, it can be verified that each tuple $(d_1,\ldots,d_k)$ in (\ref{facet_S_outerbound_}) satisfies the conditions in (\ref{ineq_G}) and (\ref{ineq_S}). 
	It follows that $\mathcal{F}_{\mathcal{S}}$ coincides with the set of tuples described by the inequalities in (\ref{facet_S_outerbound_}). 
	Hence, (\ref{facet_S_outerbound_}) is equivalent to (\ref{ineq_G}) and (\ref{ineq_S}). 
    We show the achievability of each point of $\mathcal{F}_{\mathcal{S}}$ through RS.
	First, we split $\mathcal{F}_{\mathcal{S}}$ into two subsets, denoted by $\mathcal{F}_{\mathcal{S}, 1}$  and $\mathcal{F}_{\mathcal{S},2}$, on the basis of the value of $d_{s_1}$.
    The subset $\mathcal{F}_{\mathcal{S}, 1}$ contains all the tuples of $\mathcal{F}_{\mathcal{S}}$  such that \mbox{$\alpha_{s_2} \leq d_{s_1} \leq \alpha_{s_1}$}, while $\mathcal{F}_{\mathcal{S}, 2}$ contains all the tuples of $\mathcal{F}_{\mathcal{S}}$  such that $d_{s_1} > \alpha_{s_1}$.
    So $\mathcal{F}_{\mathcal{S},1}$ is given by
	\begin{equation} 
	\begin{cases}
	\alpha_{s_2} \leq d_{s_1} \leq \alpha_{s_1} \\
	d_j \geq \alpha_j, & j \in \mathcal{S} \setminus \{s_1\} \\
	d_j \leq d_{s_1}, & j \in \bar{\mathcal{S}}_1 \\
	d_j \leq d_{s_1}, & j \in \bar{\mathcal{S}}_{21} \\
	d_j \leq \alpha_j, & j \in \bar{\mathcal{S}}_{22} \\
	d_j \leq \alpha_j, & j \in \bar{\mathcal{S}}_3 \\
	\sum_{j \in \mathcal{S}}{d_{j}}  =  1 + \sum_{j \in \mathcal{S} \setminus \{s_1\}}{\alpha_j}\\	
	\end{cases}
	\end{equation}
	where, for any value of $d_{s_1}$, the subsets $\bar{\mathcal{S}}_{21}$ and $\bar{\mathcal{S}}_{22}$ are defined as \mbox{$\bar{\mathcal{S}}_{21}=\{\, j \in \bar{\mathcal{S}}_2 \mid \alpha_j \geq d_{s_1}\,\}$} and 
	\mbox{$\bar{\mathcal{S}}_{22}=\{\, j \in \bar{\mathcal{S}}_2 \mid \alpha_j < d_{s_1}\,\}$} and they correspond to a 
	partition of $\bar{\mathcal{S}}_2$ on the basis of the value of $\alpha_j$ compared to $d_{s_1}$.
	Each admissible tuple $(d_1,\ldots,d_k)$ of $\mathcal{F}_{\mathcal{S},1}$ is achieved by RS considering $(a_1, \ldots, a_k)$ such that 
	\begin{equation}
	a_j=\begin{cases}
	  d_{s_1}, & j \in \mathcal{S} \\
   	  d_j, & j \in \bar{\mathcal{S}}_{1} \\
	  d_j, & j \in \bar{\mathcal{S}}_{21} \\
	  d_j+d_{s_1}-\alpha_j, & j \in \bar{\mathcal{S}}_{22} \\
	  d_j+d_{s_1}-\alpha_j, & j \in \bar{\mathcal{S}}_{3}.
	\end{cases}
	\end{equation}
	With such power allocation, the DoF $(d_1^{(\mathrm{p})}, \ldots, d_k^{(\mathrm{p})})$ carried by each private 
	symbol, from (\ref{DoF_private_RS}), is given by
	\begin{equation}
	d_j^{(\mathrm{p})}=\begin{cases}
    d_{s_1}, & j=s_1                     \\
    \alpha_{j}, & j \in \mathcal{S} \setminus \{s_1\}             \\
	d_j, & j \in \bar{\mathcal{S}}.
	\end{cases}
	\end{equation}
	The common symbol's DoF, which is equal to \mbox{$d^{(\mathrm{c})}=1-d_{s_1}$} from (\ref{DoF_common_symbol}), is partitioned in the following way
	 	\begin{equation}
	 	d_j^{(\mathrm{c})}=\begin{cases}
	 	0, & j=s_1 \\
	 	d_j-\alpha_j, & j \in \mathcal{S} \setminus \{s_1\} \\
	 	0, & j \in \bar{\mathcal{S}}.
	 	\end{cases}
	 	\end{equation}
   Equality in (\ref{achievable_DoF}) is satisfied and the achievability of the tuple $(d_1,\ldots,d_k)$ follows. 

    The subset $\mathcal{F}_{\mathcal{S},2}$ is equal to $\mathcal{F}_{\mathcal{S}} \setminus \mathcal{F}_{\mathcal{S},1}$  and it is given by all the non-negative real tuples $(d_1,\ldots,d_k)$ such that
	\begin{equation}
	\begin{cases}
	d_{s_1} > {\alpha}_{s_1}\\
	d_j \geq \alpha_j, & j \in \mathcal{S} \setminus \{s_1\} \\
	d_j \leq {\alpha}_{s_1}, & j \in \bar{\mathcal{S}}_1 \\
	d_j \leq \alpha_j, & j \in \bar{\mathcal{S}}_2 \\
	d_j \leq \alpha_j, & j \in \bar{\mathcal{S}}_3 \\
	\sum_{j \in \mathcal{S}}{d_{j}}  =  1 + \sum_{j \in \mathcal{S} \setminus \{s_1\}}{\alpha_j}.
	\end{cases}
	\end{equation}
	Each tuple $(d_1,\ldots,d_k)$ of $\mathcal{F}_{\mathcal{S},2}$ is achieved by RS considering $(a_1, \ldots, a_k)$ equal to
	\begin{equation}
	a_j=\begin{cases}
	\alpha_{s_1}, & j \in \mathcal{S} \\
	d_j, & j \in \bar{\mathcal{S}}_{1} \\
	d_j+\alpha_{s_1}-\alpha_j, & j \in \bar{\mathcal{S}}_{2} \\
	d_j+\alpha_{s_1}-\alpha_j, & j \in \bar{\mathcal{S}}_{3}.
	\end{cases}
	\end{equation}
    The DoF $(d_1^{(\mathrm{p})}, \ldots, d_k^{(\mathrm{p})})$ of each private
	symbol, from (\ref{DoF_private_RS}), is
	\begin{equation}
	d_j^{(\mathrm{p})}=\begin{cases}
	\alpha_{j}, & j \in \mathcal{S} \\
	d_j, & j \in \bar{\mathcal{S}}.
	\end{cases}
	\end{equation}
	The DoF carried by the common symbol, which is equal to \mbox{$d^{(\mathrm{c})}=1-{\alpha}_{s_1}$} from (\ref{DoF_common_symbol}), is partitioned in the following way
	\begin{equation}
	d_j^{(\mathrm{c})}=\begin{cases}
	d_j-\alpha_j, & j \in \mathcal{S} \\
	0, & j \in \bar{\mathcal{S}}.
	\end{cases}
	\end{equation}
	Equation (\ref{achievable_DoF}) is satisfied 
	and the tuple $(d_1, \ldots, d_k)$ is achievable. 
	Since the subsets $\mathcal{F}_{\mathcal{S},1}$ and $\mathcal{F}_{\mathcal{S},2}$ are both achievable, $\mathcal{F}_{\mathcal{S}}$ is achievable.
	Hence, the facets $\mathcal{F}_{\mathcal{S}}$ for $|\mathcal{S}| \geq 2$ are achievable.

	Next, we move to the case $|\mathcal{S}|=1$, i.e. $\mathcal{S}=\{s_1\}$.
	The set $\bar{\mathcal{S}}=\mathcal{K} \setminus \mathcal{S}$ is partitioned into two subsets, denoted as $\bar{\mathcal{S}}_1$ and $\bar{\mathcal{S}}_2$, such that 
	\mbox{$\bar{\mathcal{S}}_1=\{\, j \in \bar{\mathcal{S}} \mid j < s_1\,\}$} and  \mbox{$\bar{\mathcal{S}}_2=\{\, j \in \bar{\mathcal{S}} \mid j > s_1\,\}$}.
	In case of \mbox{$j \in \bar{\mathcal{S}}_1$}, by comparing (\ref{ineq_G}) for $\mathcal{G}=\{j, s_1 \}$ and (\ref{ineq_S}), we deduce that
    \mbox{$d_j \leq \alpha_{s_1}$}. Similarly, in case of  \mbox{$j \in \bar{\mathcal{S}}_2$}, by comparing (\ref{ineq_G}) for $\mathcal{G}=\{s_1,j \}$ and (\ref{ineq_S}), we deduce that
    \mbox{$d_j \leq \alpha_j$}.
	As earlier, $\mathcal{F}_{\mathcal{S}}$ is so rewritten as the set of 
	all the non-negative real tuples \mbox{$(d_1,\ldots,d_k)$}
	\begin{equation}
	\begin{cases}
	d_{s_1}=1 \\
	d_j \leq {\alpha}_{s_1}, & j \in \bar{\mathcal{S}}_1 \\
	d_j \leq \alpha_j, & j \in \bar{\mathcal{S}}_2. 
	\end{cases}
	\end{equation}
 	Each $(d_1,\ldots,d_k)$ is achieved by RS with $(a_1, \ldots, a_k)$
	\begin{equation}
	a_j=\begin{cases}
    \alpha_{s_1}, & j=s_1\\
	d_j, & j \in \bar{\mathcal{S}}_{1} \\
	d_j+\alpha_{s_1}-\alpha_j, & j \in \bar{\mathcal{S}}_{2}.
	\end{cases}
	\end{equation}
	The common symbol's DoF, which is equal to \mbox{$d^{(\mathrm{c})}=1-\alpha_{s_1}$}, is given to user $s_1$ only, i.e. the partition is such that \mbox{$d_{s_1}^{(\mathrm{c})}=d^{(\mathrm{c})}$} and \mbox{$d_{j}^{(\mathrm{c})}=0$ for \mbox{$j \in \mathcal{K} \setminus \{s_1\}$}}. 
    We finally consider the facets contained in the hyperplanes which delimit the half-spaces in (\ref{dis_1}).
	Taking any  $j \in \mathcal{K}$, we denote the facet contained in the hyperplane $d_j=0$ as $\mathcal{F}^{(0)}_j$. After removing the redundant inequalities, $\mathcal{F}^{(0)}_j$ is given by all the non-negative real tuples $(d_1,\ldots,d_k)$ which satisfy
	\begin{equation}
	\begin{cases}
	d_j=0 \\
    	\sum_{i \in \mathcal{S}}{d_{i}}  \leq 1 + \sum_{i \in \mathcal{S} \setminus \{s_1\}}{\alpha_i}, \quad \forall \mathcal{S} \in  \bar{\mathcal{A}}_j
	\end{cases}
	\end{equation}
	where $\bar{\mathcal{A}}_j$ is the set of all possible non-empty subsets of $\mathcal{K} \setminus \{j\}$ with elements arranged in an ascending order.
	For instance, in case of $\mathcal{K}=\{1,2,3\}$ and $j=1$, we have that \mbox{$\bar{\mathcal{A}}_j=\{\{2\},\{3\}, \{2,3\}\}$}.
	While $d_j=0$ (so user $j$ is not considered), the set of admissible tuples $(d_i)_{i \in \mathcal{K} \setminus \{j\}}$  corresponds to the region in (\ref{dis_1}) and (\ref{dis_2}) when considering the $k-1$ users $\mathcal{K}\setminus \{j\}$.
    Since we have $M$ antennas, with $M \geq k$ (hence $M$ larger than $k-1$), the facet $\mathcal{F}^{(0)}_j$ is achievable by induction hypothesis.
	Since all facets of the polyhedron are achievable, all the remaining points of the polyhedron are achievable by time-sharing. Hence, the outer-bound $\mathcal{D}$ for $K=k$ is achievable and it coincides with the \mbox{optimal DoF region $\mathcal{D}^*$}.
%
\section{Conclusion} \label{conclusion}
In this paper we show that RS is the key strategy to achieve the whole DoF region for the MISO BC with partial CSIT. 
The essence of RS, compared to conventional transmission techniques as ZFBF which rely on the transmission of private symbols only, is the transmission of a common symbol on top of the private symbols. 
The presence of the common symbol allows to tackle the multi-user interference originating from the partial CSIT more efficiently and, considering a flexible power allocation for the private symbols and flexible split of the common symbol, to achieve the entire DoF region. 
RS boils down to ZFBF in case of perfect CSIT, where the common message is not needed and ZFBF is sufficient to achieve the whole DoF region.

\ifCLASSOPTIONcaptionsoff
  \newpage
\fi

\bibliographystyle{IEEEtran}


\end{document}